\documentclass[10pt,twoside]{article}
\usepackage{graphicx}
\usepackage{psfig}
\newcommand{\be}{\begin{equation}}
\newcommand{\ee}{\end{equation}}

\makeindex

\begin{document}

\title{Star Complexes and Starburst Clumps in Spiral Galaxies}

\markboth{Yu.E.\,Efremov}{Star Complexes}

\author{Yu.N.\,Efremov$^{1}$
\\[5mm]
\it $^1$Sternberg Astronomical Institute, Moscow, Russia\\
\it e-mail: efremov@sai.msu.ru\\
}

\date{}
\maketitle

\thispagestyle{empty}

\begin{abstract}
\noindent

Star complexes are the highest level groupings in the hierarchy
of the embedded young  stars, clusters and associations,  which
obey the size - age relation. Starburst clumps,
superassociations, supergiant HII regions are different titles
for the groupings of the same size as complexes but with active
star formation over all the grouping. The coherent star formation
in the regions of violent star formation was probably triggered by
external pressure. The issue of gravitational boundness of
complexes and superassociations is briefly discussed.
\\

\noindent {\bf Keywords:} large scale star formation, clusters,
associations, star complexes, structure of galaxies
\end{abstract}

\section{Introduction}

Term  "star complexes" was suggested in 1978 to designate the
largest (up to 1 kpc)  groupings of stars and clusters which
unite objects with ages up to ~100 millions years; they were
first identified in the Sun  surroundings and one of these was
the Gould Belt system. It became soon evident that these
groupings are the same objects which have been long known as
knots and clumps in the spiral arms of external galaxies (see [1]
and references therein ).

The mechanism of coherent star formation within groupings so
large was suggested by the Elmegreens [2] who found that the
regular spacing between "the giant HII regions" (complexes) along
the spiral arms is explained by their formation from HI/CO
superclouds formed by the gravitational instability along the arm.

Star complexes are the largest and oldest objects in the
hierarchy of the young embedded groupings, which starts from the
multiple stars. The younger and smaller clusters are always
within the  older and the larger ones.  This hierarchy is similar
to one observed in the interstellar gas distribution and in fact
is the result of the latter. This similarity as well as
correlation between the ages of the oldest stars and sizes  of
groupings point to turbulence as the important factor in
formation of star groupings [3,4].

Sometimes a galaxy host one-two very bright complexes, which
consist of the OB-associations and HII regions. They were called
long ago the superassociations and represent the local
starbursts. The coherent and effective star formation in such
starburst clumps was evidently triggered by some event which has
led to the high pressure in the ISM.

\section{Star complexes in the spiral arms}

 The chains of star complexes  often delineate the spiral structure
 in the grand design galaxies.  This is observed in M31 and
 seemingly also in the Milky Way galaxies. The distribution of Cepheids
 around Sun demonstrate their strong
 concentration to the local segment of the Sgr - Car arm,
 where they form a few  vast groupings  with spacing around 1 kpc [5].
 About the same is regular spacing in the chain of the HI/CO
 superclouds along the 40 kpc long Car - Sgr arm [6].

The regular spacing  of "giant HII regions" along the arm was
noted by the Elmegreens [2] in the sample of 22 galaxies. Such a
regularity is observed in many other galaxies, but mostly in one
arm only, like it is the case for NGC 5371, NGC 4321 (M100) and
NGC 4258. This regularity mostly is not strict, but it looks like
there is a certain base spacing, some distances between complexes
being two-fold of this.

  In the galaxies with the high rate of star formation, like NGC
  6946 and M83, it is difficult to pick out the individual
 complexes from the bright arm background; in a sense, complexes
 fill  all the arm without space between them.  It is worth noting that
 the long spurs are started outward from the largest complexes
 in M31 and M51.

The regular spacing of the complexes mainly determined by joint
action of magnetic and gravitational instabilities. In NGC 6946
galaxy the  spiral arms of the regular magnetic field located
between optical ones, plausibly due to the high SFR and
disordering of magnetic lines within the arms, abundant in HII
regions [7].  The leading role of the Rayleigh-Taylor instability
in the formation of complexes is in agreement with the wave-like
appearance of some arms (including MW and M31 galaxies) in
Z-direction [8]. It is worth noting that the regular field along
the optical spiral arm is observed in M31 and MW galaxies -- and
both galaxies have the regular spacing of complexes in the long
segments of the arms.

At any rate, the regular spacing of complexes is difficult to
reconcile with  formation of superclouds in result of random
collision of smaller clouds. More so, the latter is ~3 times more
slow process than the action of the gravitational instabilities
[9].

 The short segmented arms of the flocculent spirals
 are star complexes  themselves - sheared by the differential
 galactic rotation [10].

\section{Star complexes and superassociations}

      Sometimes very bright and large blue complexes are observed
 in the arms. Many of them were noted as the superassociations (SAs),
 or starburst clumps  (see [11] and references therein).
  They are seemingly not just the extreme cases of the general distrubution
 of the stellar  complexes  in age and size (and therefore in luminosity).
 The noted examples of SAs are always much brighter than the second complex
 in a galaxy,  though in  the general distribution over a number
 of galaxies SAs might not  be divided by a gap from the lower
 luminosity complexes. This important issue is still open.

 Anyway, in the nearest SAs, 30 Dor in the LMC and OB78 in M31,
 the Cepheids and the red supergiants  are about absent.
 Thus there are no objects older than ~30 Myrs. Also, the present
 day star formation is going at low rate in 30 Dor (high density CO
 clouds are only at South-West of SA)  and more so in
 OB78, which is inside the HI superhole.

 This means that the present day  small range of ages will preserve in such SAs for ever.
 We probably observe an example of such aged SA at the East tip
 of the LMC bar, SE the cluster NGC 2065, where within region 300 pc
 in size about 180  Cepheids is observed, their range of ages
 (from the period - age relation) being  small, ~40 Myr.
 The density of Cepheids there is ~100 times higher than around Sun,
 but there is no clusters! About 100 Myr ago the region might have looked like a small
 burst of star formation. No clusters are there, like it is the case for
 OB78.

 If so, it follows that during time as long as half of the  period of the Sun
 orbital rotation compact SA  might has not dissipated yet. The strong
 concentration of complexes ~100 Myr old  within the Car -  Sgr
 spiral arm  near Sun  may implies their destruction after
 this time (due to the shearing from the galactic
 rotation  outside the arm). Otherwise  the age of the oldest
 stars in these complexes may indicate the local segment
 of the Car - Sgr arm is close to the corotation and time needed
 for the arm crossing is then quite large.

 However, the complexes formed from the virialized superclouds must be
 initially bound and then one must expect to see older complexes
 outside the arms.  After 100 Myr years  the brightest stars
 there  are too dim to permit the detection  of such complexes. Also,
 they must have been bound initially only with the parent gas, which
 since then might be blown out. The gas superclouds and star complexes
 in an arm are usually close to each other, but not coincide.

 It looks like that SAs are not just the young
 versions of the common complexes.
 Small age range implies  that star formation there was triggered by some
 source of external pressure, like the shock wave. The high pressure is favorable
 for the massive clusters to be formed bound [12].
 However, the existence of SAs without clusters suggests another
factor, such as the difference in the turbulence parameters at
the scale of the whole complex, might be in action too.
 All in all, it looks like  star groupings
 coherently formed from a supercloud might have different fates,
 depending on its location and the efficiency of star formation.

\section{Complexes of star clusters}

 The clumps of clusters have been long known in the LMC. The most
 impressive is the concentration of a dozen clusters arond NGC 2164,
  four of these  being within 200 pc and have masses around $10^5$
  sun each.  If the velocity dispersion is ~10 km/s, the virial mass of
  this cluster of clusters must be compatible with the observed mass.
  Such a group might be then  formed from the single gas supercloud,
  more so the ages of clusters there is rather similar.

  This cluster of  cluster is at the outskirts of the LMC, but
   more often the massive bound clusters  formed  within
   high pressure surroundings.  They exists, for example, in the tight clump of three rich clusters
   found recently within the brightest HII  clump in
  the young star complex NGC 5461 in the M101 galaxy [13].

   The high pressure conditions explains the numerous
   occurence of the bright  compact and presumably bound
   clusters in interacting galaxies and around centers of some spiral galaxies.
   Especially vast complexes of such clusters are known in NGC 4038/39
   (Antennae) and  are more recently discovered in NGC 7673 (Mrk 325) and
   NGC 6621/22 (see references in [11]).

   However a few such clusters
   are also known in normal spiral galaxies,
   - but always within  complexes. Two such complexes are known in
   M51. The peculiar complex in NGC 6946  is the local starburst,
   it contains  ~20 young clusters and one of them is real supercluster
   [14]. The signs of the HVC impact are observed near the cluster [15].
   This enigmatic structure is under investigation.

   It looks quite similar
   to the BCD galaxy NGC 1705, which also hosts a supercluster.
   The brightest in M101 giant HII region NGC 5471 located
outside the main body of the galaxy, it is round and has low
abundance [13]. It also strongly resembles the small BCD galaxy.
The similarities between these galaxies and the supergiant
extragalactic HII regions are well known and may be quite
significant.

The support from the grants RFBR 03-02-16288 and SciShc 389.2003.2
is appreciated.

\section*{References}

[1] Yu.N.Efremov, Astron. J., 110, 2757 (1995)

[2] B.G.Elmegreen, D.M.Elmegreen, Mon. Not., 203, 31 (1983)

[3] B.G.Elmegreen, Yu.N.Efremov, Astroph. J., 466, 802 (1996)

[4] Yu.N.Efremov, B.G.Elmegreen, Mon. Not. RAS, 299, 588 (1998)

[5] L.N.Berdnikov, Yu.N.Efremov, Astron. Lett., 19, 389, (1993)

[6] Yu.N.Efremov,  Astron. Astroph. Trans., 15, 3 (1998)

[7] P.Frick, R.Beck, A.Shukurov et al., Mon. Not. RAS, 318, 925,
(2000)

[8] J.Franco, J.Kim, E.Alfaro et al., Astroph. J., 570, 647 (2002)

[9] B.G.Elmegreen, Astroph. J., 357, 125 (1990).

 [10] B.G.Elmegreen, D.M.Elmegreen, and S.N.Leitner, Astroph. J.,
590, 271 (2003).

 [11] Yu.N.Efremov. Astrofisica, 47, 319, (2004) (in Russian).

 [12] B.G.Elmegreen, Yu.N.Efremov, Astroph. J., 480, 235 (1997)

[13] C.H.R.Chen, You-Hua-Chu, K.E.Johnson, astro-ph/0410240 (2004)

[14] S.Larsen, Yu.N.Efremov, B.G.Elmegreen et al., Astroph. J.,
567,896 (2002)

[15] Yu.N.Efremov, S.A.Pustilnik, A.Y.Kniazev et al., Astron.
Astrophys., 389, 855 (2002)

\end{document}